\documentclass[letterpaper,aps,prl,reprint,superscriptaddress,showpacs,nofootinbib,10pt,notitlepage,nobibnotes]{revtex4-2}
\usepackage{graphicx}
\usepackage{amsmath,amssymb}
\usepackage{bbm}
\usepackage[dvipsnames]{xcolor}
\usepackage[colorlinks=true,allcolors=BlueViolet,hyperfootnotes=false]{hyperref}
\hypersetup{breaklinks=true}
\usepackage[utf8]{inputenc}
\usepackage[inline]{enumitem}
\usepackage{accents}
\usepackage{stackengine}
\usepackage{orcidlink}

\usepackage{newtxtext,newtxmath}

\usepackage{tikz}

\usetikzlibrary{shapes.geometric}

\graphicspath{{Images/}}

\DeclareMathOperator\Rep{Re}
\DeclareMathOperator\Imp{Im}

\newcommand{\be}{\begin{equation}} \newcommand{\ee}{\end{equation}}
\newcommand{\ba}{\begin{array}{c}} \newcommand{\ea}{\end{array}}
\newcommand{\bea}{\begin{eqnarray}} \newcommand{\eea}{\end{eqnarray}}

\newcommand{\MeV}{\text{MeV}}
\newcommand{\fm}{\text{fm}}

\newcommand{\varlesssim}{\mathrel{\ensurestackMath{%
  \stackengine{-.4ex}{<}{\rotatebox{-25}{$\sim$}}{U}{r}{F}{T}{S}}}}
\newcommand{\vargtrsim}{\mathrel{\ensurestackMath{%
  \stackengine{-.4ex}{>}{\rotatebox{25}{$\sim$}}{U}{l}{F}{T}{S}}}}

\begin{document}

\title{\boldmath Femtoscopic signatures of the lightest $S$--wave scalar open-charm  mesons}

\newcommand{\ific}{Instituto de F\'{\i}sica Corpuscular (centro mixto CSIC-UV),
Institutos de Investigaci\'on de Paterna,
C/Catedr\'atico Jos\'e Beltr\'an 2, E-46980 Paterna, Valencia, Spain}

\newcommand{\granada}{Departamento de F\'{i}sica At\'{o}mica, Molecular y Nuclear and Instituto Carlos I de F\'{i}sica Te\'{o}rica y Computacional, Universidad de Granada, E-18071, Granada, Spain}

\author{M.~Albaladejo\orcidlink{0000-0001-7340-9235}}
\email{Miguel.Albaladejo@ific.uv.es}
\author{J.~Nieves\orcidlink{0000-0002-2518-4606}}
\email{jmnieves@ific.uv.es}
\affiliation{\ific}
\author{E.~Ruiz Arriola\orcidlink{0000-0002-9570-2552}}
\email{earriola@ugr.es}
\affiliation{\granada}

\date{\today}

\definecolor{citecolor}{rgb}{0.15,0.15,0.60}

\begin{abstract}
We predict femtoscopy correlation functions for $S$--wave $D_{(s)}\phi$ pairs of lightest pseudoscalar open charm mesons and Goldstone bosons from next-to-leading order unitarized heavy-meson chiral perturbation theory amplitudes. The effect of the two-state structure around $2300\,\MeV$ can be clearly seen in the $(S,I)=(0,1/2)$ $D\pi$, $D\eta$, $D_s \overline{K}$ correlation functions, while in the scalar-strange $(1,0)$ sector, the $D^\ast_{s0}(2317)^{\pm}$ state lying below the $DK$ threshold produces a depletion of the correlation function near threshold. These exotic states owe their existence to the nonperturbative dynamics of Goldstone-boson scattering off $D_{(s)}$. The predicted correlation functions could be experimentally measured and will shed light into the hadron spectrum confirming that it should be viewed as more than a collection of quark model states. 

\end{abstract}

\maketitle

\paragraph{Introduction.---} \label{sec:intro}

For hadrons containing charm  quarks, there is no possibility of performing traditional scattering experiments due to technical limitations imposed by the extremely short life-time of these heavy particles, and experimental information on the interactions involving charmed hadrons is commonly extracted from reactions where they are produced in the final state. In addition, lattice QCD (LQCD) is seen as a benchmark scheme to constrain and validate the effective field theories (EFTs) employed to  describe the relevant final state interactions (FSI) in the analyzed reactions. 

The study of the lightest scalar open-charm  $D^*_0(2300)$ and $D^*_{s0}(2317)^{\pm}$ provides an example of the performance of this combined strategy~\cite{Liu:2012zya,Albaladejo:2016lbb,Guo:2017jvc,Du:2017zvv,Albaladejo:2018mhb,Guo:2018kno,Sugiura:2019ane,Du:2019oki,Du:2020pui,Mai:2022eur,Asokan:2022usm}. These states are exotic in the sense that they cannot be accommodated in simple constituent quark models (see for instance the discussion in Ref.~\cite{Ortega:2016mms}). However, both resonances are dynamically generated from the S-wave scattering  of Goldstone bosons ($\phi$) off the  lightest pseudoscalar open charm  mesons [$D_{(s)}$]. The $T$--matrix is obtained after solving the Bethe-Salpeter equation with irreducible amplitudes  evaluated at next-to-leading (NLO) in  Heavy Meson Chiral Perturbation Theory (HMChPT)~\cite{Guo:2009ct,Geng:2010vw} and adopting the on-shell renormalization scheme~\cite{Oller:1997ti,Nieves:1998hp,Oller:1998zr,Nieves:1999bx,Oller:2000fj}. All the unknown low-energy constants (LECs) are determined from the LQCD $D_{(s)}\phi$ scattering lengths computed for different pion masses in Ref.~\cite{Liu:2012zya}. 
The scattering amplitudes fulfill  elastic unitarity in coupled-channels  and  successfully describe~\cite{Albaladejo:2016lbb,Albaladejo:2018mhb} the $D_{(s)}\phi$ LQCD energy levels obtained by the Hadron Spectrum~\cite{Moir:2016srx} and RQCD~\cite{Bali:2017pdv} Collaborations for various pion masses\footnote{The finite-volume scheme of Ref.~\cite{Albaladejo:2016lbb} provides predictions that also compare well with results from earlier \cite{Mohler:2013rwa,Lang:2014yfa} and more recent \cite{Gayer:2021xzv,Cheung:2020mql,Gregory:2021rgy} LQCD simulations carried out for different no-physical Goldstone boson masses.}. Moreover the theoretical framework is strongly supported~\cite{Du:2017zvv,Du:2019oki,Du:2020pui} by recent high quality $[D(\overline 
{D})\phi]$ FSI data on the $B_{(s)}\to D(\overline{D})\phi\phi^\prime$ decays provided by the LHCb experiment~\cite{LHCb:2014ioa,LHCb:2015eqv,LHCb:2015klp,LHCb:2015tsv,LHCb:2016lxy}. In the case of the $D^*_0(2300)$, the theory predicts a double-pole pattern which provides a natural explanation to various puzzles in the charm-meson spectrum~\cite{Du:2017zvv}, in particular to why a single broad $D^*_0(2300)$ resonance would have a mass almost equal or even higher than its strange sibling $D^*_{s0}(2317)^{\pm}$. Actually, it was only in the 2018 edition of the Review of Particle Physics (RPP)~\cite{ParticleDataGroup:2018ovx} when the possible existence of a lighter state associate to the $D^*_0(2300)$ was first suggested based on the findings of Refs.~\cite{Albaladejo:2016lbb,Du:2017zvv,Guo:2018kno}.

On the other hand, femtoscopy has been traditionally utilized to measure the size of the quark-gluon plasma fireball created in relativistic heavy-ion collisions. However, since it is sensitive to the correlations between the particles in the final state, the parameters of the strong interaction can be probed as well~\cite{ALICE:2020mfd,ALICE:2022wwr}. In high-multiplicity events of $pp$, $pA$ and $AA$ collisions, the hadron production yields are well described by the statistical models, thereby leaving the correlations between outgoing particles relying upon the FSI,  from which the corresponding scattering parameters can be extracted~\cite{Fabbietti:2020bfg}. Femtoscopy techniques consist in measuring the correlation in momentum space for any hadron-hadron pair, and this information is encoded in the so-called two-particle correlation function (CF). The latter can be computed as the quotient of the number of pairs of combined particles with the same relative momentum produced in the same collision event over the reference distribution of pairs from mixed events. On the theoretical side, CF comes in terms of the product of the so-called source function, which can be seen as a kind of form factor that gives the probability that the two particles forming a pair are emitted at a relative distance of each other,  times the squared of the absolute value of the wave function of the considered channel, built from the scattering amplitudes~\cite{Koonin:1977fh, Lednicky:1981su,Pratt:1986cc,Pratt:1990zq,Bauer:1992ffu,Morita:2014kza, Ohnishi:2016elb,Morita:2016auo, Hatsuda:2017uxk, Mihaylov:2018rva,Haidenbauer:2018jvl,Morita:2019rph, Kamiya:2019uiw,Kamiya:2021hdb,Kamiya:2022thy, Vidana:2023olz,Liu:2023uly}. 

Additional independent information on the charm-hadron spectrum would be most welcome to further learn about its properties and nature. Femtoscopic CFs with the observation of channels involving charmed hadrons in heavy ion collisions should
be then most valuable. There exist experimental studies in the strangeness sector~\cite{ALICE:2018ysd,ALICE:2019gcn,ALICE:2021njx,ALICE:2019buq,ALICE:2019eol,ALICE:2019hdt,ALICE:2020mfd,ALICE:2021cpv,ALICE:2021cyj}, but importantly the ALICE collaboration measurement of the $pD^-$ and $\overline{p}D^+$ CFs in high-multiplicity $pp$ collisions at 13 TeV \cite{ALICE:2022enj} paves the way to access the charm quark sector. This is precisely the goal of this letter, where we will predict the visible signatures that the $D^*_{s0}(2317)^{\pm}$ and the two-pole structure of the $D^*_0(2300)$ will produce in the $D^0 K^+$, $D^+ K^0$ and $D_s^+ \eta$, and
the $D^+\pi^0$, $D^0\pi^+$ $D^+\eta$ and $D_s^+ \overline{K}^0$ CFs, respectively. None of these channels involve the Coulomb interaction and therefore the unitarized NLO HMChPT scheme, constrained and tested by data and LQCD input, should perform very well and provide very accurate results which could be confronted to experiment or used to validate femtoscopic techniques and models.

\paragraph{Correlation functions.---} \label{sec:formalism}

The fundamental quantities of our study are the CFs, which can be written as: 
\begin{subequations}\label{eq:CFs}
\begin{equation}\label{eq:CFsGen}
    C_i(s) = 1 + \int_{0}^\infty \!\!\!\!\! \mathrm{d}r\, S(r) 
    \Big(
    \sum_{j}
    \left\lvert
    \psi_{i}^{\,j}(s,r)
    \right\rvert^2
    - j_0 \left( p_i\, r \right)^2
    \Big)\,,
    \end{equation}\vspace{-12pt}
    \begin{equation}
 \psi_{i}^{\,j}(s,r) = j_0 \left( p_i\, r \right)\delta_{ij} + T_{ji
 }(s) \widetilde{G}_j(s,r)\,,
\end{equation}
\end{subequations}
where the usual Mandelstam variable $s$ denotes the c.m. energy squared. The index $i$ denotes the coupled hadron pair considered, and the sum in the index $j$ runs over all possible ones. Here we consider the channels $D\pi$, $D\eta$, and $D_s \overline{K}$ in the $(S,I)=(0,1/2)$ sector , and $D_s^+ \pi^0$, $D^0 K^+$, $D^+ K^0$, and $D_s^+ \eta$  for strangeness $S=1$ and isospin $I=0$ and $I=1$. The c.m.~momentum of each channel is denoted by $p_i(s)$, and $j_0(x)=\sin(x)/x$ is the zeroth order spherical Bessel function. The functions $\widetilde{G}_i(s,r)$, given in Ref.\,\cite{Vidana:2023olz}, can be written as:
\begin{equation}\label{eq:Gtilde}
    \widetilde{G}_i(s,r) = \frac{1}{\pi} \int_{s_{\text{th},i}}^{\infty}\!\!\!\! \mathrm{d}s'\,
    \frac{p_i(s')}{8\pi\sqrt{s'}}
    \,
    \frac{j_0(p_i(s')\,r)}{s-s'+i\epsilon}\,\theta\left(\Lambda - p_i(s') \right).
\end{equation}
The sharp momentum cutoff $\Lambda$ is introduced to regularize the $r \to 0$ behaviour of the wave functions $\psi_i^{\,j}(s,r)$. In the work of Ref.~\cite{Vidana:2023olz}, it is inherited from the regularization of the amplitude $T(s)$, and allows the on-shell factorization of the latter out of the integral. As it will be shown, its effect is very small for natural $\Lambda$ values in the range $\left[ 600,900 \right]\,\MeV$.

The source function is generally taken as a Gaussian, which in the radial direction reads $S(r) = 4\pi r^2/(4\pi R^2)^{3/2} \exp(-r^2/4R^2)$, with a source size $R$. A factor $4\pi r^2$ coming from the differential volume element $\mathrm{d}^3 \vec{r}$ has been absorbed into $S(r)$.

\begin{figure}
    \centering
    \includegraphics[width=8.75cm]{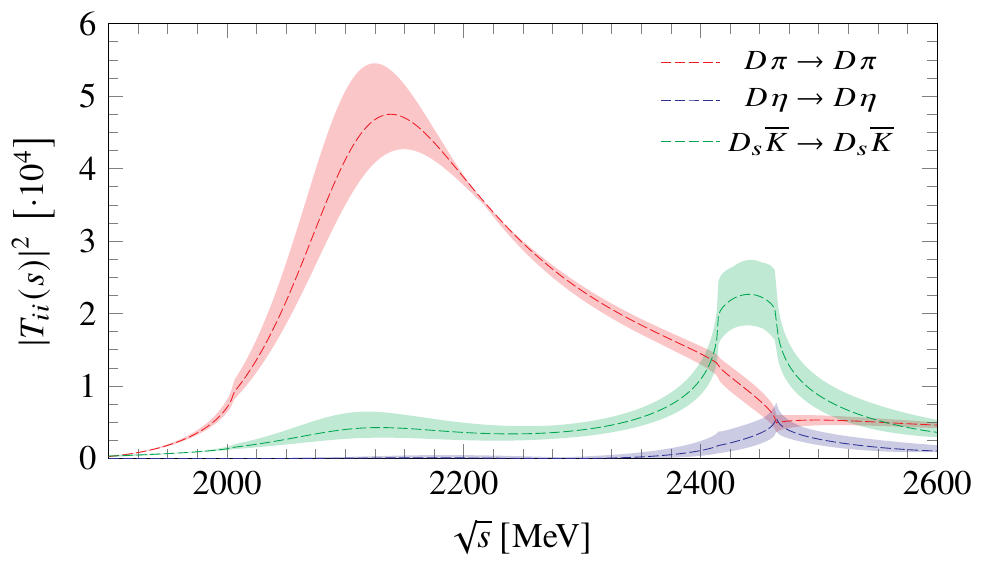}
    \caption{Modulus squared of the diagonal elements of the $I=1/2$ $J^P = 0^+$ $T$-matrix in Eq.~\eqref{eq:TMat}. \label{fig:Amplitudes}}
\end{figure}

The key ingredient in the CFs are the the $S$-wave unitarized partial wave amplitudes $T_{ij}(s)$, which encode the meson--meson interactions. In this  work, they are taken from Ref.~\cite{Liu:2012zya} and written as:
\begin{equation}\label{eq:TMat}
    T^{-1}(s) = V^{-1}(s) - G(s)~,
\end{equation}
with elementary interactions $V(s)$ computed from NLO HMChPT, and $G(s)$ are two-meson loop functions regularized with a subtraction constant. As mentioned in the Introduction, the LECs as well as the subtraction constants have been successfully determined in Ref.\,\cite{Liu:2012zya} through the computation of $D_{(s)}\phi$ scattering lengths for different pion masses. Therefore, these amplitudes are completely determined (no free parameters) and, in this sense, the calculations presented in this manuscript are genuine predictions for future experiments. The amplitudes, in turn, have been used in Refs.\,\cite{Albaladejo:2016lbb} and \cite{Albaladejo:2018mhb} to compute energy levels in a finite volume for the $D\pi$, $D\eta$, and $D_s\overline{K}$  and  $DK$ systems, that are successfully compared with the LQCD simulations in Refs.~\cite{Moir:2016srx} and \cite{Bali:2017pdv}, respectively. Importantly, in Ref.~\cite{Albaladejo:2016lbb} it is shown that these amplitudes contain two poles in two different Riemann sheets, corresponding to two different $D^\ast_0$ states in the $2300\,\MeV$ region, and not just one (previously known as $D^\ast_0(2400)$, see Refs.~\cite{ParticleDataGroup:2018ovx,ParticleDataGroup:2022pth}), as traditionally thought. Later on, Ref.~\cite{Du:2017zvv} showed that this two-state structure is also compatible with the experimental LHCb data  on $B^- \to D^+ \pi^- \pi^-$~\cite{LHCb:2016lxy} and $B^0_s \to \overline{D}^{\,0} K^- \pi^+$~\cite{LHCb:2014ioa}. As already mentioned, this two-state picture has been further tested using FSI data from LHCb experiment, and is now considered as a strong possibility in the current edition of the RPP \cite{ParticleDataGroup:2022pth}. For later reference, the modulus squared of the diagonal matrix elements are shown in Fig.~\ref{fig:Amplitudes}, where a clear peak around $2135\,\MeV$ produced by the lightest $D^\ast_0$ state can be seen, and the structure of the second resonance can be appreciated also at $2450\,\MeV$, together with a strong interference with the $D\eta$ and $D_s \overline{K}$ thresholds.

\begin{figure*}[t]
    \centering
    \includegraphics[width=8.75cm]{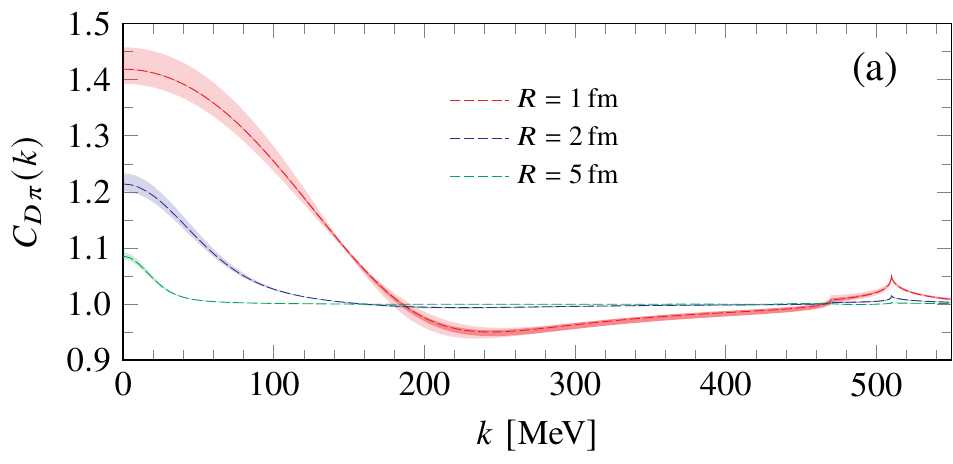}
    \includegraphics[width=8.75cm]{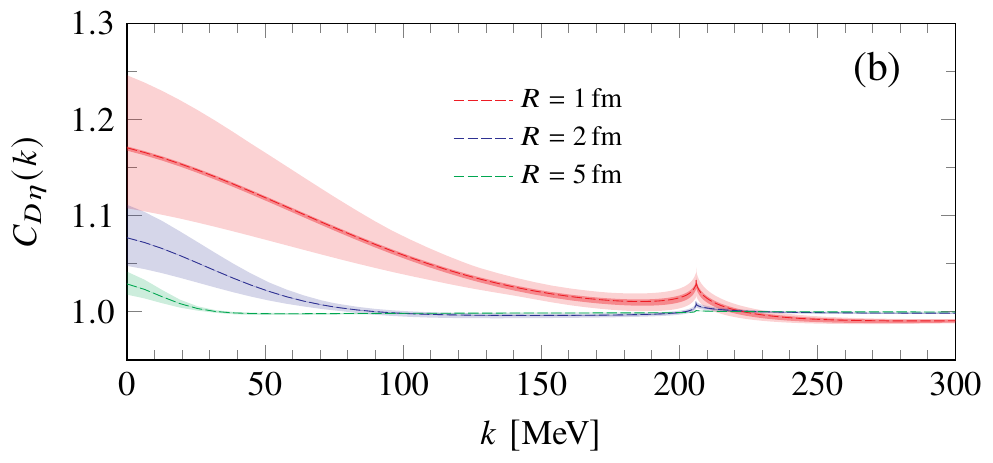}
    \includegraphics[width=8.75cm]{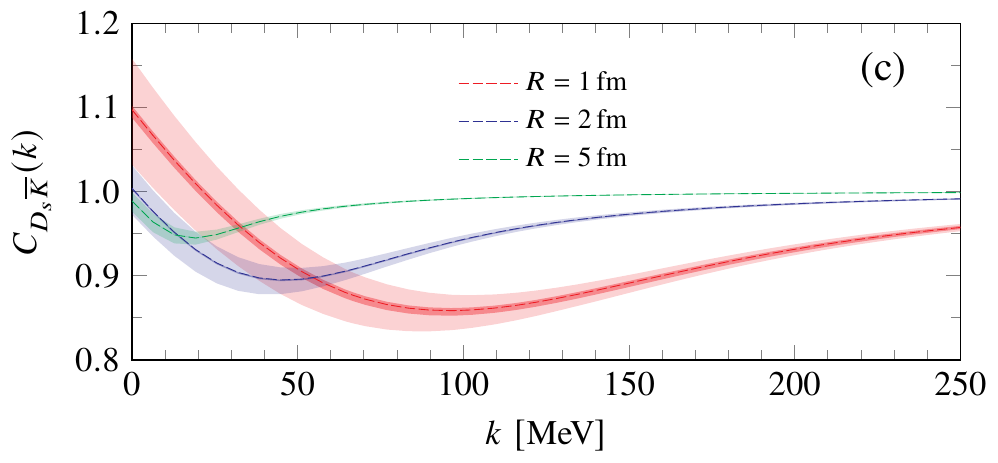}
    \includegraphics[width=8.75cm]{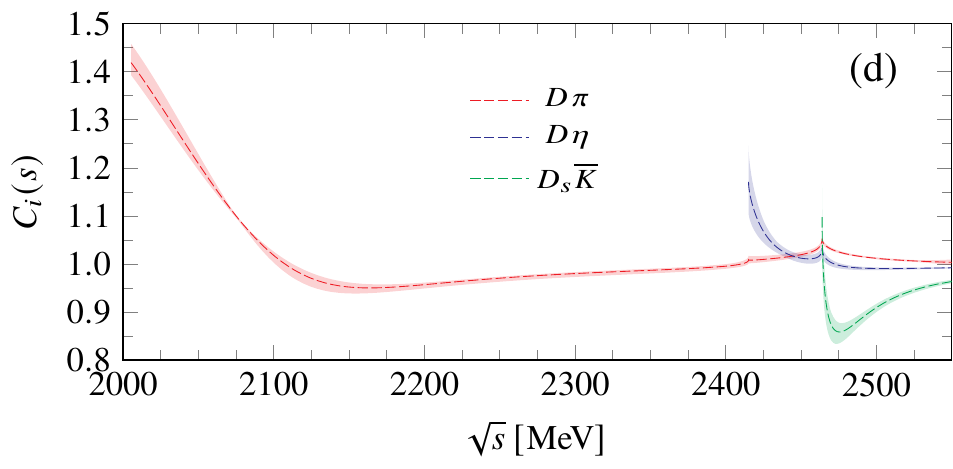}
    \caption{Correlation function for the $D\pi$ (a), $D\eta$ (b), and $D_s \overline{K}$ (c) channels with $I=I_z=1/2$ as a function of their c.m. momentum $k$ for different values of the source size $R=1\,\fm$ (red), $2\,\fm$ (blue), and $5\,\fm$. The dark, inner bands represent the variation of $\Lambda$ in Eq.~\eqref{eq:Gtilde} in a range $\left[600,900\right]\,\MeV$. The light, outer bands represent the addition in quadrature of the 68\% confident-level (CL) uncertainties in the CFs inherited from those affecting  the LECs of the NLO unitarized scattering amplitudes. In panel (d) we present together the CFs, as a function of the c.m. energy $\sqrt{s}$, for the three channels with a source of size $R=1\,\fm$, and with only the full uncertainty band.\label{fig:TPCF}}
\end{figure*}

\paragraph{Non-strange sector and the two $D^{\ast}_0$ states.---}\label{sec:isovector}

The CFs for the $I=1/2$ channels are shown in Fig.~\ref{fig:TPCF} for different, typical sizes $R=1$, $2$, and $5\,\fm$ of the source. One would expect the clear peak at $\sqrt{s}\simeq 2135\,\MeV$ in the $D\pi$ amplitude (see Fig.~\ref{fig:Amplitudes}), produced by the lightest $D^\ast_0$ state, to leave a strong imprint in the $C_{D\pi}(k)$ CF at $k \simeq 215\,\MeV$. Instead of a bump, a depletion and a minimum can be found there for the $R=1\,\fm$ source, which might come as a surprise. However, one can show (see Appendix) from the Lednicky-Lyuboshits model~\cite{Lednicky:1981su} that this result can be expected for a Breit--Wigner-like resonance with a width $\Gamma \simeq 200\,\MeV$ and for a source size $R \vargtrsim 1\,\fm$. For $R=2$ and $5\,\fm$, the minimum is still present, but certainly diluted. One would need a much narrower state ($\Gamma \varlesssim 50\,\MeV$) or smaller source size ($R \varlesssim 0.5\,\fm$) to observe a clear peak. Nonetheless, the position of the minimum closer to $215\,\MeV$ rather than to $400\,\MeV$ (as expected for a nominal mass around $2340\,\MeV$ \cite{ParticleDataGroup:2022pth}) would be a clear indication of the existence of a lower pole.

The strongest effect of the second (higher) pole can be seen in the $D_s \overline{K}$ amplitude depicted in Fig.\,\ref{fig:Amplitudes}, although it is located below the threshold. Indeed, one can observe in $C_{D_s \overline{K}}(k)$ [Fig.~\ref{fig:TPCF}, panel (c)] a clear dip for low momentum $k$, \textit{i.e.} for energies $\sqrt{s}$ slightly above the $D_s\overline{K}$ threshold. 

All these features can be better appreciated in Fig.~\ref{fig:TPCF}, panel (d), where the CFs of the three pairs are shown for $R=1\,\fm$ as a function of $\sqrt{s}$. There, the two different minima at $\sqrt{s} \simeq 2135\,\MeV$ ($D\pi$ CF) and $2475\,\MeV$ ($D_s\overline{K}$ CF), produced by the two different $D^\ast_0$ states, can be observed. The experimental observation of these features would constitute a strong additional support of the two-state pattern predicted by the HMChPT unitary amplitudes employed.

Finally, we note that we have considered the CF for the $I=1/2$ $D\pi$ system, which, for $I_z = +1/2$, is a linear combination of the physical $D^+ \pi^0$ and $D^0 \pi^+$.\footnote{Note that, by considering $I_z=+1/2$, we avoid the interference with Coulomb interaction. The latter is present in the case $I_z=-1/2$, which mixes $D^+ \pi^-$ and $D^0 \pi^0$, and would be also present in $D_s^+ K^-$ channel. Note also that the isospin breaking effect between the $D^+ \pi^0$ and $D^0\pi^+$ states induced by the masses is expected to be small, since their thresholds are just $1\,\MeV$ apart.} To obtain the truly measurable CFs for the physical $D\pi$ states, one would need to consider also the $I=3/2$ components. Introducing the appropriate isospin combinations in Eq.~\eqref{eq:CFs} one finds:
\begin{equation}
C_{D^+ \pi^0} = \frac{2}{3} C^{D\pi}_{3/2} + \frac{1}{3} C^{D\pi}_{1/2}~, \quad
C_{D^0 \pi^+} = \frac{1}{3} C^{D\pi}_{3/2} + \frac{2}{3} C^{D\pi}_{1/2}~.
\end{equation}
However, since the interaction in $I=3/2$ is very weak, one finds (as we have checked) $C^{D\pi}_{3/2} \simeq 1$. Therefore, both physical CFs can be obtained from our computed $C^{D\pi}_{1/2}(k)$ displayed in Fig.~\ref{fig:TPCF}, and in particular we also predict the relation $2\,C_{D^0\pi^+}(k)-C_{D^+\pi^0}(k) \simeq 1$.

\begin{figure}[t]
    \centering
    \includegraphics[width=8.75cm]{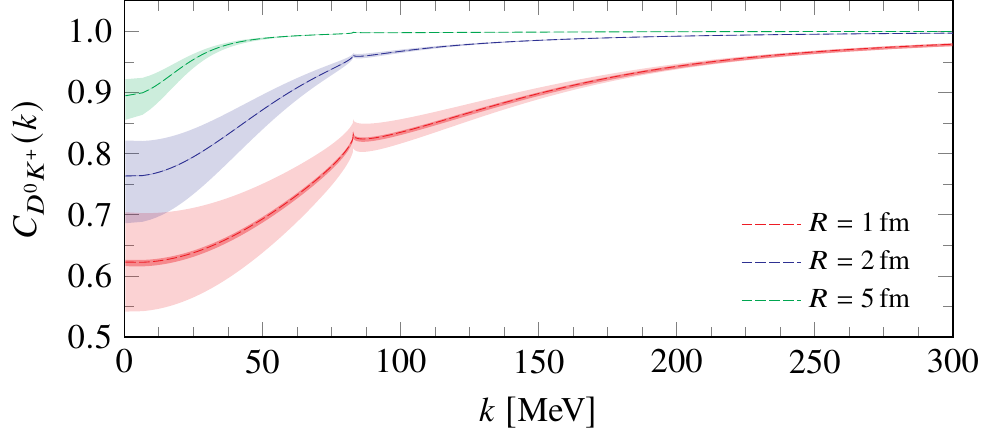}
    \caption{$D^0 K^+$ correlation function for different source sizes, $R=1\,\fm$ (red), $2\,\fm$ (blue), and $5\,\fm$.  \label{fig:Ds2317}}
\end{figure}

\paragraph{Strange sector and the $D^\ast_{s0}(2317)^\pm$ state.---} In the $S=1$ sector we consider the $D_s^+ \pi^0$, $D^0 K^+$, $D^+ K^0$, and $D_s^+ \eta$ channels, with $I=0$ and $I=1$ components. The CF for the $D^0 K^+$ channel (the lowest lying of the two $DK$ states with $I_z=0$) is shown in Fig.~\ref{fig:Ds2317} for different source sizes $R$. The $D^+ K^0$ threshold lies $\simeq 9\,\MeV$ above the $D^0 K^+$ one, which implies a momentum $k_{D^0 K^+} \simeq 83\,\MeV$, where a cusp effect can be seen in Fig.~\ref{fig:Ds2317}. Furthermore, and unlike the $D\pi$ CFs mentioned before, the $DK$ CFs cannot be filtered, because:
\begin{equation}
    C_{D^0 K^+} = C_{D^+ K^0} = \frac{C^{DK}_{0} + C^{DK}_{1}}{2}\,
\end{equation}
with $C^{DK}_{I}$ the CFs for definite isospin $I$. This fact, together with the larger mass difference between the $D^0 K^+$ and $D^+ K^0$ thresholds, prompts one to use physical channels from the beginning instead of definite isospin ones\footnote{The interaction in the $DK$ isovector channel, though smaller than in the scalar sector which  generates the exotic $D^\ast_{s0}(2317)^\pm$, is not negligible in contrast to the case of the $I=3/2$ $D\pi$ channel.}.

A clear depletion in $C_{D^0 K^+}$ at the origin can be seen in Fig.~\ref{fig:Ds2317}, being the suppression larger for the smaller source sizes. This effect is due to the presence in our amplitudes $T_{ij}(s)$ of the $D^\ast_{s0}(2317)^\pm$ bound-state pole\footnote{In the isospin limit the $I=0$ and $I=1$ sectors decouple, and the $D^{\ast}_{s0}(2317)^\pm$ is a bound state in the $I=0$ amplitudes below the $DK$ threshold. This state lies above the $I=1$ $D_s^+\pi^0$ channel but has a negligible coupling to it when the latter is coupled to the physical $DK$ ones. In this sense, we still call the $D^\ast_{s0}(2317)^\pm$ a bound state.} below the $DK$ threshold, with a binding energy around $45\,\MeV$, which translates into a large binding momentum $p=i\gamma_b$, $\gamma_b \simeq 190\,\MeV$. The observation of this behavior at the $D^0 K^+$ threshold in the measurement of these CFs, and in particular their dependence on the source size $R$, could provide important feedback on the exotic state $D^\ast_{s0}(2317)^{\pm}$.

The $D^0 K^+$ CF has been recently computed in Ref.~\cite{Liu:2023uly}. There, the chiral amplitudes were computed only at LO and the $T_{ij}$ amplitudes were regularized with Gaussian-like form factors, inducing a different off-shell behavior of the $T$--matrix which also produces variations in the wave-functions. The cutoff was fixed by adjusting the position of the generated pole to the mass of the $D^\ast_{s0}(2317)^{\pm}$, but this does not guaranty that the $DK$ $I=1/2$ scattering length obtained in Ref.~\cite{Liu:2023uly} is equal to that calculated in our scheme. These differences could explain the changes that can be seen in the CFs, in particular at $k\simeq 0$ where, for instance, the scheme of Ref.~\cite{Liu:2023uly} predicts $C_{D^0 K^+}(0) \simeq 0.4$ for $R=1.2\,\fm$. However, the trend is similar for both Ref.~\cite{Liu:2023uly} and the present work, which is reassuring. Moreover, the CF uncertainty bands in our results might partially account for the variations.

\paragraph{Conclusions.---} We have presented the first joint prediction of the CFs of the open-charm $D_{(s)}\phi$ pairs in the $(S,I)=(0,1/2)$ and $S=1$ sectors based on well determined NLO unitarized chiral amplitudes. For $I=1/2$ two distinct minima are observed in the correlation functions $C_{D\pi}$ and $C_{D_s \overline{K}}$, which are produced by the lower and higher $D^\ast_0$ poles, respectively. The measurement of these features of the CFs could shed light on this otherwise elusive resonance(s). We also predict $2\,C_{D^0\pi^+}(k) - C_{D^+\pi^0}(k) \simeq 1$.
For the $D^0 K^+$ pair, a sizeable depletion is observed at threshold in the CF (as in Ref.~\cite{Liu:2023uly}), due to the presence of the $D^\ast_{s0}(2317)^{\pm}$ bound state originated from the $I=0$ interaction. 

Given the expected precision that can be achieved in the measurements of CFs at ALICE, and other experiments that could join the femtoscopy venture, a clear comparison with the CFs predicted in this work should be feasible in a near future. In particular, the observation of the features produced by the two-state pattern in the $D\pi$ and $D_s \overline{K}$ CFs would constitute a further strong support of this picture. In addition to the experimental consequences of the measurements of these correlation functions, the comparison with the predictions presented in this work would certainly help to reduce the theoretical uncertainties of the chiral amplitudes.  

\paragraph{Acknowledgements.---} The authors would like to thank Ot\'on V\'azquez Doce for valuable discussions. This work was supported by the Spanish Ministerio de Ciencia e Innovaci\'on (MICINN) under contracts Nos.\,PID2020-112777GB-I00 and PID2020-114767GB-I00, by Generalitat Valenciana under contract PROMETEO/2020/023 and Junta de Andaluc\'ia grant FQM-225. This project has received funding from the European Union Horizon 2020 research and innovation programme under the program H2020-INFRAIA-2018-1, grant agreement No.\,824093 of the STRONG-2020 project.  M.\,A.~is supported through Generalitat Valenciana (GVA) Grants No.\,CIDEGENT/2020/002 and thanks the warm support of ACVJLI. 

\appendix
\section{Lednicky-Lyuboshits model and resonances}\label{app:LLsimplemodel}
In the Lednicky-Lyuboshits approximation for a single-channel CF \cite{Lednicky:1981su}, one substitutes the full wave function $\psi(s,r)$ in Eq.~\eqref{eq:CFsGen} with its non-relativistic, asymptotic ($r\to \infty$) form:
\begin{equation}\label{eq:psiasy}
    \phi_\text{asy}(k,r) \simeq \frac{ \sin{kr}}{k r} + f(k) \frac{e^{ikr}}{r}\,,
\end{equation}
where for clarity the momentum $k$ instead of the Mandelstam variable $s$ has been taken as a variable. Introducing Eq.~\eqref{eq:psiasy} into Eq.~\eqref{eq:CFsGen}, and assuming still a gaussian source $S(r)$, the resulting CF is:
\begin{equation}\label{eq:CLL}
    C_\text{LL}(k) = 1 
    + \frac{\left\lvert f(k) \right\rvert^2}{2R^2} 
    + \frac{2\Rep{f(k)}}{\sqrt{\pi}R} F_1(x)
    - \frac{\Imp{f(k)}}{R} F_2(x)~,
\end{equation}
where $x=2kR$, $F_1(x) = \int_{0}^{x} \mathrm{d}t e^{(t^2 - x^2)}/x$, and $F_2(x) = \left( 1 - e^{-x^2} \right)/x$. Above $f(k)$ is the standard quantum mechanics amplitude, $\Imp{f(k)}=-k$. 
It can be written as:
\begin{equation}\label{eq:fk}
    f^{-1}(k) = k \cot\delta(k) - i k~,
\end{equation}
where $\delta(k)$ is the phase shift, and one gets:
\begin{equation}\label{eq:CLLcot}
    C_\text{LL}(k) = 1 + 
    \frac{2\sin^2\delta(k)}{x^2} 
    \left( 
    e^{-x^2} + \frac{2x F_1(x)}{\sqrt{\pi}} \cot\delta(k)
    \right)\,.
\end{equation}
At threshold,
\begin{equation}
C_\text{LL}(k=0) = 1 + \frac{a\left(a + \frac{4R}{\sqrt{\pi}}\right)}{2R^2}\,,
\end{equation}
where $a$ is the scattering length, $a = \lim_{k \to 0} k \cot{\delta(k)}$. For $a>0$ or $a < -\frac{4}{\sqrt{\pi}} R \simeq -2.26\,R$, one has $C_\text{LL}(k=0)>1$. The lowest value $C_\text{LL}(k=0)$ can achieve is for $a=-\frac{2}{\sqrt{\pi}}R \simeq -1.13\,R$, when one has $C_\text{LL}(k=0) = 1-\frac{2}{\pi} \simeq 0.36$. Incidentally, the $I=0$ $DK$ scattering length is $a=-0.84\,\fm$, thus not far from this value for $R=1\,\fm$. The depletion at threshold in the $D^0 K^+$ CF is thus caused by its $I=0$ component. 

On a different note, if there is a resonance for a momentum $k_R$, then $\delta(k_R)=\pi/2$, and one has $C_\text{LL}(k_R) = 1 +  \frac{2e^{-x_R^2}}{x_R^2}$, with $x_R = 2 k_R R$. Therefore, for large $x_R$, say $x_R \gtrsim 2$, the correlation function is largely suppressed, $C_\text{LL}(k_R) \simeq 1$, slightly above unity. Furthermore, with a regular Breit-Wigner shape for the phase shift $\delta(k)$, it can also be proven that $C'(k_R) < 0$, \textit{i.e.} the CF is decreasing at $k=k_R$. For $k > k_R$ one has $\cot\delta(k)<0$ ($\delta(k)>\pi/2$), which is a necessary condition for the bracket in Eq.~\eqref{eq:CLLcot} to change sign. It is then possible for $C_\text{LL}(k)$ to go below one for $k>k_R$, specially taking into account that the first term in the bracket is very suppressed with respect to the coefficient  of $\cot\delta(k)$ in the second term. Since asymptotically for large momenta one has $C_\text{LL}(k) \to 1$, then a minimum will appear if the CF becomes smaller than one for some momentum region above $k_R$.
\begin{figure}
    \centering
    \includegraphics[width=8.5cm]{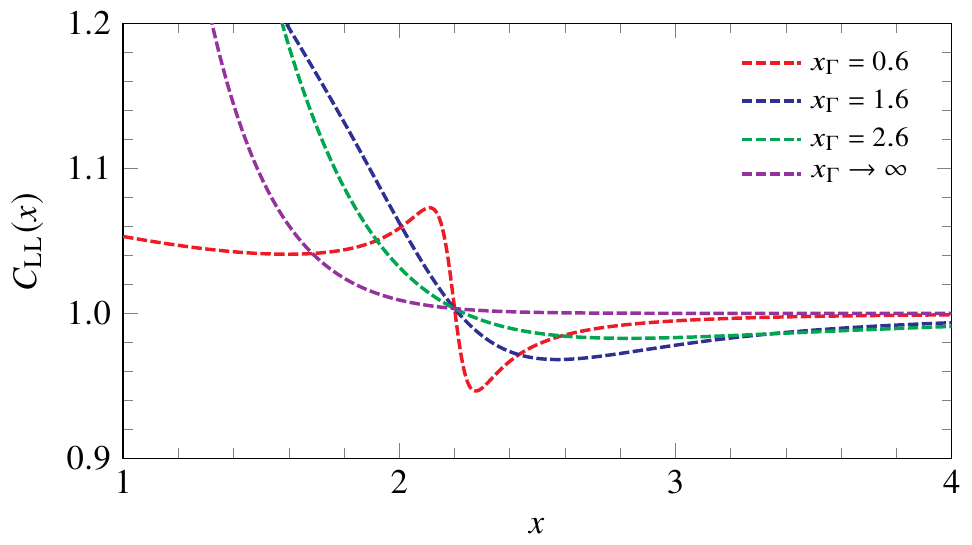}
    \caption{Correlation function in the Lednicky-Lyuboshits approximation ($C_\text{LL}(k)$), in terms of the reduced variable $x = 2kR$, for different values of the reduced width $x_\Gamma^2 = 4\mu \Gamma R^2$ (see text for further details). \label{fig:SimpleModelLL}}
\end{figure}

Certainly, whether this minimum is pronounced or diluted, or a maximum can appear for $k < k_R$, depends on the details of the resonance ($k_R$ and its width, $\Gamma$) and the source size $R$. For illustrative purposes, let us assume a non-relativistic BW-shape for $\cot\delta(k)$,
\begin{equation}
    \cot\delta(k) = -\frac{k_R}{k}\frac{k^2 - k_R^2}{2\mu}\frac{2}{\Gamma} = -\frac{x_R}{x}\frac{x^2 - x_R^2}{x_\Gamma^2}\,,
\end{equation}
with $x_\Gamma^2 = 4\,\mu\,\Gamma\,R^2$, and $\mu$ the reduced mass of the hadron pair. We show in Fig.~\ref{fig:SimpleModelLL} $C_\text{LL}(k)$ (as a function of the reduced variable $x$) obtained for different values of $x_\Gamma$, taking, for definiteness, $x_R = 2.2$ (for $R=1\,\fm$, this would correspond to $k_R \simeq 215\,\MeV$, and $\sqrt{s} = 2105\,\MeV$ for a $D\pi$ system). As discussed earlier, since $x_R \gtrsim 2$, all the curves are very close to one and decreasing around $x \simeq x_R$. As can be seen, for a narrow resonance ($x_\Gamma = 0.6$, red dashed line) there appears a maximum before $x_R$, but there is still a minimum after. In the case of a broader resonance with $x_\Gamma = 1.8$ (blue dashed line) (which would correspond to $\Gamma \simeq 200\,\MeV$ for $R=1\,\fm$ and $D\pi$ reduced mass $\mu$) only a minimum can be appreciated. For larger $x_\Gamma$, the minimum can no longer be observed.

\bibliographystyle{apsrev4-1}
\bibliography{references.bib}

\newpage

\end{document}